\documentclass[dvips,sissa]{prl}

\begin{document}

\newcommand{\noi}{\noindent}
\newcommand{\ie}{\textit{i.e.,\xspace}}
\newcommand{\eg}{\textit{e.g.,\xspace}}
\newcommand{\etc}{\textit{etc.\xspace}}
\newcommand{\cf}{\textit{cf.\xspace}}
\newcommand{\etal}{\textit{et al.\xspace}}
\newcommand{\half}{\frac{1}{2}}
\newcommand{\third}{\frac{1}{3}}
\newcommand{\fourth}{\frac{1}{4}}
\newcommand{\sss}{\scriptscriptstyle}

\renewcommand{\Im}{\text{Im}}
\renewcommand{\Re}{\text{Re}}
\renewcommand{\vec}[1]{\ensuremath{\mathbf{#1}}}


\newcommand{\prfo}{Pr$^{\text{IV}}$\xspace}
\newcommand{\prth}{Pr$^{\text{III}}$\xspace}
\newcommand{\cuth}{Cu$^{\text{III}}$\xspace}
\newcommand{\refo}{R$^{\text{IV}}$\xspace}
\newcommand{\htc}{high-$T_c$\xspace}
\newcommand{\tjm}{$t$-$J$ model\xspace}
\newcommand{\tjmsl}{t-J model\xspace}


\newcommand{\ybco}{YBa$_2$Cu$_3$O$_{7-\delta}$\xspace}
\newcommand{\pbco}{PrBa$_2$Cu$_3$O$_{7-\delta}$\xspace}
\newcommand{\srcocl}{Sr$_2$CuO$_2$Cl$_2$\xspace}
\newcommand{\pbcosl}{PrBa$_{\text{2}}$Cu$_{\text{3}}$O$_{\text{7-$\delta$}}$\xspace}
\newcommand{\prox}{Pr-O$_7$\xspace}
\newcommand{\yox}{Y-O$_7$\xspace}
\newcommand{\yprox}{Y$_{1-x}$Pr$_x$-O$_7$\xspace}
\newcommand{\rprox}{R$_{1-x}$Pr$_x$-O$_7$\xspace}
\newcommand{\cuo}{CuO$_2$\xspace}
\newcommand{\cuot}{CuO$_3$\xspace}


\renewcommand{\k}{\ensuremath{\mathbf{k}}\xspace}
\renewcommand{\r}{\ensuremath{\mathbf{r}}\xspace}


\newcommand{\ghat}{\ensuremath{\widehat g}\xspace}
\newcommand{\Ghat}{\ensuremath{\widehat G}\xspace}
\newcommand{\Shat}{\ensuremath{\widehat \Sigma}\xspace}
\newcommand{\That}{\ensuremath{\widehat T}\xspace}
\newcommand{\shat}{\ensuremath{\widehat\sigma}\xspace}


\newcommand{\wt}{\widetilde\omega\xspace}
\newcommand{\dkt}{\widetilde\Delta_{\k}\xspace}
\newcommand{\xkt}{\widetilde\xi_{\k}\xspace}


\newcommand{\dw}{\ensuremath{d_{x^2 - y^2}}\xspace}
\newcommand{\textdw}{
  \ensuremath{\text{d}_{\text{x}^{\text{2}} - \text{y}^{\text{2}}}}\xspace}
\newcommand{\tc}{\ensuremath{T_c}\xspace}
\newcommand{\ef}{\ensuremath{\epsilon_F}\xspace}
\newcommand{\wn}{\ensuremath{\omega_n}\xspace}
\newcommand{\wnt}{\ensuremath{\widetilde \omega_n}\xspace}
\newcommand{\dk}{\ensuremath{\Delta_{\k}}\xspace}
\newcommand{\dsh}{\ensuremath{\Delta_{\text{sh}}}\xspace}
\newcommand{\xk}{\ensuremath{\xi_{\k}}\xspace}
\newcommand{\nres}{\ensuremath{N_{\text{res}}}\xspace}
\newcommand{\akw}{\ensuremath{A(\k, \omega)}\xspace}
\newcommand{\frsdes}{\ensuremath{{}_{\vphantom{f}\sss f}^{\vphantom{l} \sss 2} 
               \pi_{\vphantom{l}s}^{\phantom{\dagger}}}\xspace}
\newcommand{\fcr}{\ensuremath{{}^\nu\!\! f_{\vphantom{l}ms}^\dagger}\xspace}
\newcommand{\fdes}{\ensuremath{{}^\nu\!\! f_{\vphantom{l}ms}^{
\phantom{\dagger}}}\xspace}
\newcommand{\cs}[1]{\ensuremath{c_{#1\sigma}}}
\newcommand{\csd}[1]{\ensuremath{c_{#1\sigma}^\dagger}}
\newcommand{\csdd}[1]{\ensuremath{c_{#1\downarrow}^\dagger}}
\newcommand{\csdu}[1]{\ensuremath{c_{#1\uparrow}^\dagger}}

\newcommand{\sumijs}{\ensuremath{\sum_{\langle i,j \rangle, \sigma}}}  
\newcommand{\sumij}{\ensuremath{\sum_{\langle i,j \rangle}}}
\newcommand{\llvert}{\left\vert}
\newcommand{\rrvert}{\right\vert}
\newcommand{\llangle}{\left\langle}
\newcommand{\rlangle}{\right\langle}
\newcommand{\lrangle}{\left\rangle}
\newcommand{\rrangle}{\right\rangle}

\begin{multicols}{2}

\noindent
{\large\bf
Comment on ``Single Hole Dynamics in the \cuo Plane at Half Filling''
}

\medskip

In a recent letter \cite{pothuizen-etal-97}, Pothuizen \etal reported
angle-resolved photoemission (ARPES) data on the single-layer insulating
cuprate \srcocl. They observed a strongly \k dependent intensity: closest to
\ef, a weak feature near ($\pi/2, \pi/2$) (assigned to the Zhang-Rice singlet
state \cite{zhang-rice-88}, consistent with Ref. \cite{wells-etal-95}), and at
high-symmetry points of the Brillouin zone and lower energy, relatively sharp
peaks with large intensity. The spectra were analyzed in terms of a nearest
neighbour (nn) tight-binding model using all O 2$p$ and Cu 3$d$
orbitals. From this, it was concluded that the strong intensity at the
high-symmetry points comes from {\em purely oxygen-like} orbitals which due to
the lack of 3$d$ hybridization are very much single-particle-like. We argue
that an assignment of the intense features to pure oxygen states is too
simple, and inconsistent with LDA calculations for \srcocl \cite{jepsen-97},
which are expected to give accurate results for the single-particle properties
we are interested in here.

An interpretation of the intense peaks in terms of pure oxygen states is
{\em only} possible at the zone corner $\k = (\pi, \pi)$ for the state in
Fig. 1c. It has $g$ symmetry with respect to the Cu site, and therefore, cannot
mix with any Cu 3$d$ or Cl 3$p$ levels. Furthermore, the LDA results
\cite{jepsen-97} show that the admixture of Sr 3$d$ levels is also
neglibible because of their large energy separation from the O 2$p$
states.

However, the situation is entirely different at $\k = (0,0)$ and $\k =
(\pi,0)$: In both cases, the O states used in the analysis of
Ref. \cite{pothuizen-etal-97} can hybridize effectively with Cl 3$p_{x,y}$
levels as shown in Figs. 1a,b. This is confirmed by LDA \cite{jepsen-97} which
shows similar admixtures of O 2$p$ and Cl 3$p$ levels for the states closest to
\ef at those \k points. Consequently, it is not possible to extract a {\em
generic} (usable as a microscopic model parameter for {\em all} cuprates) nn
O-O hopping matrix element $t_{pp}$ from the energy difference of the intense
peaks at the different high-symmetry points as proposed in
\cite{pothuizen-etal-97}. There are too many additional {\em material-specific}
parameters (at least the O-Cl hopping matrix element, and the O-Cl on-site
energy difference) which enter the expression for the energy splittings, not
just $t_{pp}$.

Nevertheless, we agree with the authors of \cite{pothuizen-etal-97} that the
strong intensity of these peaks is most likely due to the absence of mixing
with the strongly correlated $d$ states. This interpretation is further
suggested by the fact that at $\k = (\pi,0)$ the LDA predicts the topmost band
below \ef to have strong Cu 3$d_{x^2 - y^2}$ admixture, which might explain why
the corresponding feature in the ARPES is not very sharp.

Another point concerns the width of the intense peaks: The LDA results show
that at none of the \k points considered is there a {\em single} well separated
band with no $d$ character just below \ef. There are always at least two bands
with planar character, split by an amount comparable or less than 400meV, the
approximate experimental width of the sharp peaks. Hence, their spectral weight
is possibly a superposition from more than one band. 
This might also be partially responsible for their large intensity.
Furthermore, a substantial part of the line-width broadening is due to
momentum resolution. The size of this contribution depends on the
dispersion, and is therefore generally different at different \k points
\cite{fehrenbacher-96c}. In any case, conclusions about scattering mechanisms
from a comparison of line-widths as attempted in \cite{pothuizen-etal-97} have
to be taken with caution.

\vspace{3mm}
\begin{center}
\includegraphics[width=6.0cm]{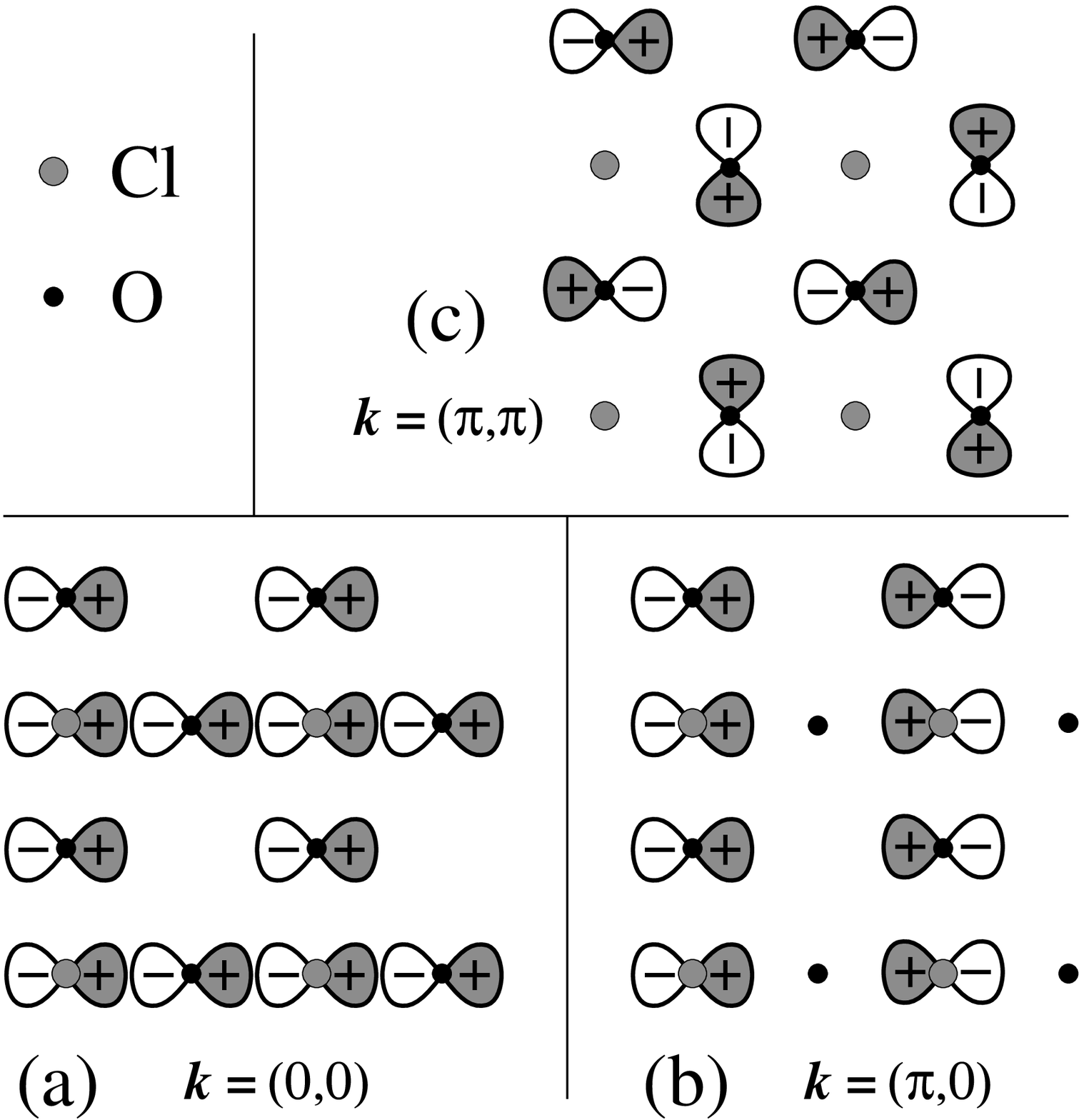}
\mycap{Possible combinations of Oxygen 2$p$ and Chlorine 3$p$ orbitals at
high-symmetry points of the Brillouin zone, (a) $\k = (0,0)$, (b) $\k = (\pi,
0)$, (c) $\k = (\pi, \pi)$.
}
\vspace{1mm}
\end{center}

Finally, we would like to mention that LDA indeed predicts that the state
closest to \ef at $\k = (\pi, \pi)$ is the one depicted in Fig. 1c (almost
degenerate with a mixed Cl 3$p_z$, Cu 3$d_{3z^2 - r^2}$ level), as proposed in
\cite{pothuizen-etal-97}. The energy difference of this state as compared to
the Zhang-Rice singlet feature is crucial in determining the stablity
of the \prfo oxidation state, which is thought to be responsible for the \tc
suppression in \pbco \cite{fehrenbacher-rice-93}. The experiments in
\cite{pothuizen-etal-97} therefore impose some experimental constraints for
the possible parameters of the model proposed in
\cite{fehrenbacher-rice-93}. We shall elaborate on this point in future work.

We acknowledge useful discussions with G. Stollhoff, and O. Jepsen, and wish to
thank O. Jepsen for providing his LDA results to us prior to publication.

\medskip

{\small
\noindent R. Fehrenbacher
\renewcommand{\theenumi}{}
\renewcommand{\labelenumi}{\theenumi}
\vspace{-1.3mm}
\begin{enumerate}
\item
Max-Planck-Institut f\"ur Festk\"orper\-forschung \\
Heisenbergstr. 1 \\
D-70569 Stuttgart \\
Germany
\end{enumerate}
}

\end{multicols}


\vspace{-1mm}


\begin{thebibliography}{1}

\bibitem{pothuizen-etal-97}
J.~J.~M. Pothuizen et~al.,
\newblock Phys. Rev. Lett. {\bf 78}, 717 (1997)\relax
\relax
\bibitem{zhang-rice-88}
F.~C. Zhang and T.~M. Rice,
\newblock Phys. Rev. B {\bf 37}, 3759 (1988)\relax
\relax
\bibitem{wells-etal-95}
B.~O. Wells et~al.,
\newblock Phys. Rev. Lett. {\bf 74}, 964 (1995)\relax
\relax
\bibitem{jepsen-97}
O.~Jepsen,
\newblock unpublished\relax
\relax
\bibitem{fehrenbacher-96c}
R.~Fehrenbacher,
\newblock Phys. Rev. B {\bf 54}, 6632 (1996)\relax
\relax
\bibitem{fehrenbacher-rice-93}
R.~Fehrenbacher and T.~M. Rice,
\newblock Phys. Rev. Lett. {\bf 70}, 3471 (1993)\relax
\relax
\end{thebibliography}
\end{document}